\documentclass[conference]{IEEEtran}

\usepackage{cite}
\usepackage[T1]{fontenc} 
\ifCLASSINFOpdf
\else
\fi
\usepackage{bm,color,soul}
\usepackage{amsmath, amsthm, amssymb}
\theoremstyle{plain}

\usepackage[pdftex]{graphics}
\theoremstyle{definition}

\usepackage{xspace}
\usepackage[nolist,printonlyused]{acronym}    
\usepackage{amssymb}
\usepackage{graphicx}
\usepackage{amsmath}
\usepackage{mathtools}
\usepackage{graphicx,color} 
\usepackage{amsmath, amsfonts, amssymb, amsbsy,nccmath} 
\usepackage{algorithm} 
\usepackage{enumerate} 
\usepackage{algorithmic} 
\usepackage{lipsum} 
\usepackage{epstopdf}
\usepackage{mathtools}
\usepackage{dsfont} 
\usepackage[inline]{enumitem}
\usepackage{amsmath}
\usepackage[cmintegrals]{newtxmath}
\usepackage{color,soul}
\usepackage{stfloats}  
\usepackage{tabularx}
\usepackage{lipsum}
\usepackage{mwe}
\usepackage{amsmath}
\usepackage{longtable}
\usepackage{subcaption}

\usepackage[T1]{fontenc}
\usepackage[utf8]{inputenc}
\usepackage{authblk}

\DeclareCaptionFont{xipt}{\fontsize{8}{13}\mdseries}
\usepackage[font=xipt]{caption}

\usepackage[T1]{fontenc}
\usepackage[utf8]{inputenc}
\usepackage{authblk}

\begin{document}  \pagestyle{empty}

\title{Adaptive Timers and Buffer Optimization for Layer-2 Protocols in 5G Non-Terrestrial Networks}

  \pagestyle{empty}


\vspace{-3mm}
\author{Chandan~Kumar~Sheemar, Sumit Kumar,
Jorge Querol, 
and Symeon Chatzinotas\\  
Interdisciplinary Centre for Security, Reliability and Trust (SnT), University of Luxembourg, Luxembourg\\
emails: \{chandankumar.sheemar sumit.kumar  jorge.querol  symeon.chatzinotas\}@uni.lu
}

\maketitle
\begin{acronym}
  \acro{2G}{Second Generation}
  \acro{3G}{3$^\text{rd}$~Generation}
  \acro{3GPP}{3$^\text{rd}$~Generation Partnership Project}
  \acro{4G}{4$^\text{th}$~Generation}
  \acro{5G}{5$^\text{th}$~Generation}
  \acro{AA}{Antenna Array}
  \acro{AC}{Admission Control}
  \acro{ACL}{adjacent channel leakage}
  \acro{AD}{Attack-Decay}
  \acro{ADC}{analog-to-digital converter}
  \acro{ADSL}{Asymmetric Digital Subscriber Line}
  \acro{AHW}{Alternate Hop-and-Wait}
  \acro{AMC}{Adaptive Modulation and Coding}
	\acro{AP}{Access Point}
  \acro{APA}{Adaptive Power Allocation}
  \acro{AR}{autoregressive}
  \acro{ARMA}{Autoregressive Moving Average}
  \acro{ATES}{Adaptive Throughput-based Efficiency-Satisfaction Trade-Off}
  \acro{AWGN}{additive white Gaussian noise}
  \acro{A/D}{analog and digital}
  \acro{BB}{branch and bound}
  \acro{BD}{block diagonalization}
  \acro{BER}{bit error rate}
  \acro{BF}{Best Fit}
  \acro{BLER}{BLock Error Rate}
  \acro{BPC}{Binary power control}
  \acro{BPSK}{binary phase-shift keying}
  \acro{BPA}{Best \ac{PDPR} Algorithm}
  \acro{BRA}{Balanced Random Allocation}
  \acro{BS}{base station}
  \acro{CAP}{Combinatorial Allocation Problem}
  \acro{CAPEX}{Capital Expenditure}
  \acro{CBF}{Coordinated Beamforming}
  \acro{CBR}{Constant Bit Rate}
  \acro{CBS}{Class Based Scheduling}
  \acro{CC}{Congestion Control}
  \acro{CDF}{Cumulative Distribution Function}
  \acro{CDMA}{Code-Division Multiple Access}
  \acro{CL}{Closed Loop}
  \acro{CLI}{cross-link interference}
  \acro{CLPC}{Closed Loop Power Control}
  \acro{CNR}{Channel-to-Noise Ratio}
  \acro{CPA}{Cellular Protection Algorithm}
  \acro{CPICH}{Common Pilot Channel}
  \acro{CoMP}{Coordinated Multi-Point}
  \acro{CQI}{Channel Quality Indicator}
  \acro{CRM}{Constrained Rate Maximization}
	\acro{CRN}{Cognitive Radio Network}
  \acro{CS}{Coordinated Scheduling}
  \acro{CSI}{channel state information}
  \acro{CSIR}{channel state information at the receiver}
  \acro{CSIT}{channel state information at the transmitter}
  \acro{CUE}{cellular user equipment}
  \acro{DAC}{digital-to-analog converter}
  \acro{D2D}{device-to-device}
  \acro{DCA}{Dynamic Channel Allocation}
  \acro{DE}{Differential Evolution}
  \acro{DFT}{Discrete Fourier Transform}
  \acro{DIST}{Distance}
  \acro{DL}{downlink}
  \acro{DMA}{Double Moving Average}
	\acro{DMRS}{Demodulation Reference Signal}
  \acro{D2DM}{D2D Mode}
  \acro{DMS}{D2D Mode Selection}
  \acro{DOCSIS}{Data Over Cable Service Interface Specification}
  \acro{DPC}{Dirty Paper Coding}
  \acro{DRA}{Dynamic Resource Assignment}
  \acro{DSA}{Dynamic Spectrum Access}
  \acro{DSM}{Delay-based Satisfaction Maximization}
  \acro{EBD}{electrical balance duplexer}
  \acro{ECC}{Electronic Communications Committee}
  \acro{EFLC}{Error Feedback Based Load Control}
  \acro{EI}{Efficiency Indicator}
  \acro{eNB}{Evolved Node B}
  \acro{EPA}{Equal Power Allocation}
  \acro{EPC}{Evolved Packet Core}
  \acro{EPS}{Evolved Packet System}
  \acro{E-UTRAN}{Evolved Universal Terrestrial Radio Access Network}
  \acro{ES}{Exhaustive Search}
  \acro{FD}{full-duplex}
  \acro{FDD}{frequency division duplexing}
  \acro{FDM}{Frequency Division Multiplexing}
  \acro{FER}{Frame Erasure Rate}
  \acro{FF}{Fast Fading}
  \acro{FSB}{Fixed Switched Beamforming}
  \acro{FST}{Fixed SNR Target}
  \acro{FTP}{File Transfer Protocol}
  \acro{GA}{Genetic Algorithm}
  \acro{GBR}{Guaranteed Bit Rate}
  \acro{GLR}{Gain to Leakage Ratio}
  \acro{GOS}{Generated Orthogonal Sequence}
  \acro{GPL}{GNU General Public License}
  \acro{GRP}{Grouping}
  \acro{HARQ}{Hybrid Automatic Repeat Request}
  \acro{HD}{half-duplex}
  \acro{HMS}{Harmonic Mode Selection}
  \acro{HOL}{Head Of Line}
  \acro{HSDPA}{High-Speed Downlink Packet Access}
  \acro{HSPA}{High Speed Packet Access}
  \acro{HTTP}{HyperText Transfer Protocol}
  \acro{ICMP}{Internet Control Message Protocol}
  \acro{ICI}{Intercell Interference}
  \acro{ID}{Identification}
  \acro{IETF}{Internet Engineering Task Force}
  \acro{ILP}{Integer Linear Program}
  \acro{JRAPAP}{Joint RB Assignment and Power Allocation Problem}
  \acro{UID}{Unique Identification}
  \acro{IBFD}{in-band full-duplex}
  \acro{IID}{Independent and Identically Distributed}
  \acro{IIR}{Infinite Impulse Response}
  \acro{ILP}{Integer Linear Problem}
  \acro{IMT}{International Mobile Telecommunications}
  \acro{INV}{Inverted Norm-based Grouping}
  \acro{IoT}{Internet of Things}
  \acro{IP}{Internet Protocol}
  \acro{IPv6}{Internet Protocol Version 6}
  \acro{I/Q}{in-phase/quadrature}
  \acro{IRS}{Intelligent Reflective Surface}
  \acro{ISAC}{integrated sensing and communications}
  \acro{ISD}{Inter-Site Distance}
  \acro{ISI}{Inter Symbol Interference}
  \acro{ITU}{International Telecommunication Union}
  \acro{JCAS}{joint communications and sensing}
  \acro{JOAS}{Joint Opportunistic Assignment and Scheduling}
  \acro{JOS}{Joint Opportunistic Scheduling}
  \acro{JP}{Joint Processing}
	\acro{JS}{Jump-Stay}
    \acro{KF}{Kalman filter}
  \acro{KKT}{Karush-Kuhn-Tucker}
  \acro{L3}{Layer-3}
  \acro{LAC}{Link Admission Control}
  \acro{LA}{Link Adaptation}
  \acro{LC}{Load Control}
  \acro{LNA}{low-noise amplifier}
  \acro{LOS}{Line of Sight}
  \acro{LP}{Linear Programming}
  \acro{LS}{least squares}
  \acro{LTE}{Long Term Evolution}
  \acro{LTE-A}{LTE-Advanced}
  \acro{LTE-Advanced}{Long Term Evolution Advanced}
  \acro{M2M}{Machine-to-Machine}
  \acro{MAC}{Medium Access Control}
  \acro{MANET}{Mobile Ad hoc Network}
  \acro{MC}{Modular Clock}
  \acro{MCS}{Modulation and Coding Scheme}
  \acro{MDB}{Measured Delay Based}
  \acro{MDI}{Minimum D2D Interference}
  \acro{MF}{Matched Filter}
  \acro{MG}{Maximum Gain}
  \acro{MH}{Multi-Hop}
  \acro{MIMO}{multiple-input multiple-output}
  \acro{MINLP}{Mixed Integer Nonlinear Programming}
  \acro{MIP}{Mixed Integer Programming}
  \acro{MISO}{Multiple Input Single Output}
  \acro{ML}{maximum likelihood}
  \acro{MLWDF}{Modified Largest Weighted Delay First}
  \acro{mmWave}{millimeter wave}
  \acro{MME}{Mobility Management Entity}
  \acro{MMSE}{minimum mean squared error}
  \acro{MOS}{Mean Opinion Score}
  \acro{MPF}{Multicarrier Proportional Fair}
  \acro{MRA}{Maximum Rate Allocation}
  \acro{MR}{Maximum Rate}
  \acro{MRC}{maximum ratio combining}
  \acro{MRT}{Maximum Ratio Transmission}
  \acro{MRUS}{Maximum Rate with User Satisfaction}
  \acro{MS}{mobile station}
  \acro{MSE}{mean squared error}
  \acro{MSI}{Multi-Stream Interference}
  \acro{MTC}{Machine-Type Communication}
  \acro{MTSI}{Multimedia Telephony Services over IMS}
  \acro{MTSM}{Modified Throughput-based Satisfaction Maximization}
  \acro{MU-MIMO}{multiuser multiple input multiple output}
  \acro{MU}{multi-user}
  \acro{NAS}{Non-Access Stratum}
  \acro{NB}{Node B}
  \acro{NE}{Nash equilibrium}
  \acro{NCL}{Neighbor Cell List}
  \acro{NLP}{Nonlinear Programming}
  \acro{NLOS}{Non-Line of Sight}
  \acro{NMSE}{Normalized Mean Square Error}
  \acro{NORM}{Normalized Projection-based Grouping}
  \acro{NP}{Non-Polynomial Time}
  \acro{NR}{New Radio}
  \acro{NRT}{Non-Real Time}
  \acro{NSPS}{National Security and Public Safety Services}
  \acro{O2I}{Outdoor to Indoor}
  \acro{OFDMA}{orthogonal frequency division multiple access}
  \acro{OFDM}{orthogonal frequency division multiplexing}
  \acro{OFPC}{Open Loop with Fractional Path Loss Compensation}
  \acro{O2I}{Outdoor-to-Indoor}
  \acro{OL}{Open Loop}
  \acro{OLPC}{Open-Loop Power Control}
  \acro{OL-PC}{Open-Loop Power Control}
  \acro{OPEX}{Operational Expenditure}
  \acro{ORB}{Orthogonal Random Beamforming}
  \acro{JO-PF}{Joint Opportunistic Proportional Fair}
  \acro{OSI}{Open Systems Interconnection}
  \acro{PAIR}{D2D Pair Gain-based Grouping}
  \acro{PAPR}{Peak-to-Average Power Ratio}
  \acro{PBCH}{physical broadcast channel}
  \acro{P2P}{Peer-to-Peer}
  \acro{PC}{Power Control}
  \acro{PCI}{Physical Cell ID}
  \acro{PDF}{Probability Density Function}
  \acro{PDPR}{pilot-to-data power ratio}
  \acro{PER}{Packet Error Rate}
  \acro{PF}{Proportional Fair}
  \acro{P-GW}{Packet Data Network Gateway}
  \acro{PL}{Pathloss}
  \acro{PMN}{Perceptive Mobile Network}
  \acro{PPR}{pilot power ratio}
  \acro{PRB}{physical resource block}
  \acro{PROJ}{Projection-based Grouping}
  \acro{ProSe}{Proximity Services}
  \acro{PS}{Packet Scheduling}
  \acro{PSAM}{pilot symbol assisted modulation}
  \acro{PSO}{Particle Swarm Optimization}
  \acro{PZF}{Projected Zero-Forcing}
  \acro{QAM}{Quadrature Amplitude Modulation}
  \acro{QoS}{Quality of Service}
  \acro{QPSK}{Quadri-Phase Shift Keying}
  \acro{RAISES}{Reallocation-based Assignment for Improved Spectral Efficiency and Satisfaction}
  \acro{RAN}{Radio Access Network}
  \acro{RA}{Resource Allocation}
  \acro{RAT}{Radio Access Technology}
  \acro{RATE}{Rate-based}
  \acro{RB}{resource block}
  \acro{RBG}{Resource Block Group}
  \acro{REF}{Reference Grouping}
  \acro{RF}{radio frequency}
  \acro{RIS}{reconfigurable intelligent surface}
  \acro{RLC}{Radio Link Control}
  \acro{RM}{Rate Maximization}
  \acro{RNC}{Radio Network Controller}
  \acro{RND}{Random Grouping}
  \acro{RRA}{Radio Resource Allocation}
  \acro{RRM}{Radio Resource Management}
  \acro{RSCP}{Received Signal Code Power}
  \acro{RSRP}{Reference Signal Receive Power}
  \acro{RSRQ}{Reference Signal Receive Quality}
  \acro{RR}{Round Robin}
  \acro{RRC}{Radio Resource Control}
  \acro{RSSI}{Received Signal Strength Indicator}
  \acro{RT}{Real Time}
  \acro{RU}{Resource Unit}
  \acro{RUNE}{RUdimentary Network Emulator}
  \acro{RV}{Random Variable}
  \acro{RX}{receiver}
  \acro{SAC}{Session Admission Control}
  \acro{SBFD}{sub-band full-duplex}
  \acro{SCM}{Spatial Channel Model}
  \acro{SC-FDMA}{Single Carrier - Frequency Division Multiple Access}
  \acro{SD}{Soft Dropping}
  \acro{S-D}{Source-Destination}
  \acro{SDPC}{Soft Dropping Power Control}
  \acro{SDMA}{Space-Division Multiple Access}
  \acro{SER}{Symbol Error Rate}
  \acro{SES}{Simple Exponential Smoothing}
  \acro{S-GW}{Serving Gateway}
  \acro{SI}{self-interference}
  \acro{SINR}{signal-to-interference-plus-noise ratio}
  \acro{SIC}{self-interference cancellation}
  \acro{SIP}{Session Initiation Protocol}
  \acro{SISO}{single-input single-output}
  \acro{SIMO}{Single Input Multiple Output}
  \acro{SIR}{signal-to-interference ratio}
  \acro{SLNR}{Signal-to-Leakage-plus-Noise Ratio}
  \acro{SMA}{Simple Moving Average}
  \acro{SNR}{signal-to-noise ratio}
  \acro{SORA}{Satisfaction Oriented Resource Allocation}
  \acro{SORA-NRT}{Satisfaction-Oriented Resource Allocation for Non-Real Time Services}
  \acro{SORA-RT}{Satisfaction-Oriented Resource Allocation for Real Time Services}
  \acro{SPF}{Single-Carrier Proportional Fair}
  \acro{SRA}{Sequential Removal Algorithm}
  \acro{SRS}{Sounding Reference Signal}
  \acro{STAR}{simultaneous transmit-and-receive}
  \acro{SU-MIMO}{single-user multiple input multiple output}
  \acro{SU}{Single-User}
  \acro{SVD}{Singular Value Decomposition}
  \acro{TCP}{Transmission Control Protocol}
  \acro{TDD}{time division duplexing}
  \acro{TDMA}{Time Division Multiple Access}
  \acro{TETRA}{Terrestrial Trunked Radio}
  \acro{TP}{Transmit Power}
  \acro{TPC}{Transmit Power Control}
  \acro{TTI}{Transmission Time Interval}
  \acro{TTR}{Time-To-Rendezvous}
  \acro{TSM}{Throughput-based Satisfaction Maximization}
  \acro{TU}{Typical Urban}
  \acro{TX}{transmit}
  \acro{RX}{receive}
  \acro{UE}{User Equipment}
  \acro{UEPS}{Urgency and Efficiency-based Packet Scheduling}
  \acro{UL}{uplink}
  \acro{UMTS}{Universal Mobile Telecommunications System}
  \acro{URI}{Uniform Resource Identifier}
  \acro{URM}{Unconstrained Rate Maximization}
  \acro{UT}{user terminal}
  \acro{VR}{Virtual Resource}
  \acro{VoIP}{Voice over IP}
  \acro{WAN}{Wireless Access Network}
  \acro{WCDMA}{Wideband Code Division Multiple Access}
  \acro{WF}{Water-filling}
  \acro{WiMAX}{Worldwide Interoperability for Microwave Access}
  \acro{WINNER}{Wireless World Initiative New Radio}
  \acro{WLAN}{Wireless Local Area Network}
  \acro{WMPF}{Weighted Multicarrier Proportional Fair}
  \acro{WPF}{Weighted Proportional Fair}
  \acro{WSN}{Wireless Sensor Network}
  \acro{WWW}{World Wide Web}
  \acro{XIXO}{(Single or Multiple) Input (Single or Multiple) Output}
  \acro{XDD}{cross-division duplex}
  \acro{ZF}{zero-forcing}
  \acro{ZMCSCG}{Zero Mean Circularly Symmetric Complex Gaussian}
\end{acronym}

\begin{abstract}
   Interest in the integration of Terrestrial Networks (TN) and Non-Terrestrial Networks (NTN); primarily satellites; has been rekindled due to the potential of NTN to provide ubiquitous coverage. Especially with the peculiar and flexible physical layer properties of 5G-NR, now direct access to 5G services through satellites could become possible. However, the large Round-Trip Delays (RTD) in NTNs require a re-evaluation of the design of RLC and PDCP layers timers ( and associated buffers), in particular for the regenerative payload satellites which have limited computational resources, and hence need to be optimally utilized. Our aim in this work is to initiate a new line of research for emerging NTNs with limited resources from a higher-layer perspective. To this end, we propose a novel and efficient method for optimally designing the RLC and PDCP layers' buffers and timers without the need for intensive computations. This approach is relevant for low-cost satellites, which have limited computational and energy resources. The simulation results show that the proposed methods can significantly improve the performance in terms of resource utilization and delays. 
\end{abstract}
 \pagestyle{empty}
 
\IEEEpeerreviewmaketitle
 \pagestyle{empty}
\section{Introduction}
Non-Terrestrial Networks (NTNs) refer to communication networks that operate in space, beyond the boundaries of Earth's surface \cite{kodheli2017integration}. These networks utilize satellites, space-based platforms, or other orbiting infrastructure to establish connectivity and enable communication services. NTNs play a crucial role in bridging the digital divide, providing global coverage, and extending connectivity to remote and un-derserved areas where terrestrial infrastructure is limited or absent. They are instrumental in supporting various applications, including telecommunications, internet access, broadcasting, and disaster response \cite{kodheli2020satellite}, by leveraging the advantages of satellite-based communication systems in terms of wide coverage, mobility, and scalability \cite{giordani2020satellite}.
 
  \pagestyle{empty}
 However, compared to conventional terrestrial networks (TNs) \cite{sheemar2021game,sheemar2022practical,sheemar2021beamforming,sheemar2023full,sheemar2022near}, NTNs suffer additional challenges due to extremely large round-trip delays (RTDs) for geosynchronous Earth orbits (GEO), medium Earth orbit (MEO), and low Earth orbit (LEO) satellites\cite{hoyhtya2022sustainable}.
 However, the benefits of NTNs have motivated 3GPP, which has been involved in the standardization of TN communications only \cite{sheemar2021massive,thomas2019multi,sheemar2020receiver}, to take a big leap toward studying the issues and providing solutions to integrate NTN in the 5G ecosystem. From 3GPP Release 17 onward, satellites have become an integral part of 5G deployment \cite{38811, 38821}, and this integration is coined as 5G-NTN. 
 
 In 3GPP Release 17, a particular emphasis was given to the transparent payload, with the objective of providing direct access to the 5G services to user equipments (UEs) on Earth. Transparent payload-based satellites do not require expensive hardware. Their main goal is to amplify and forward the signals coming from a ground gNB to a ground UE and vice versa, acting like a high-altitude relay, with the potential to establish communication between two entities located even very far on Earth. This is a simple setup that enables the reuse of the existing fleet of satellites while both the gNB and the UE operate on the ground.

\begin{figure}
\includegraphics[width=\linewidth]{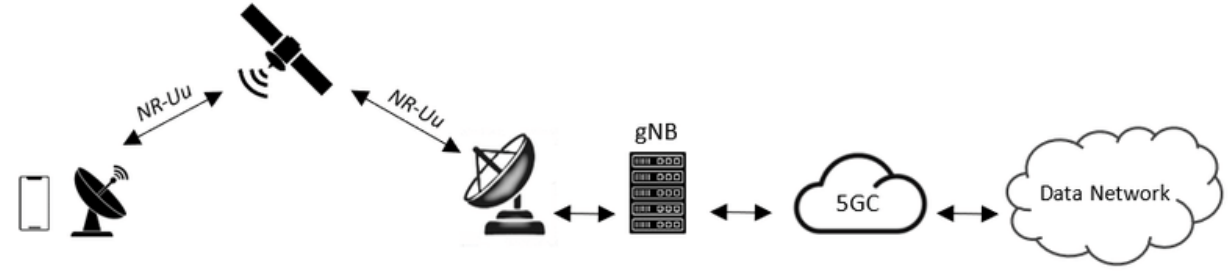}
   \caption{Fifth generation non-terrestrial network with transparent payload. }
   \label{fig_Ng1} 
\end{figure}

\begin{figure}
    \includegraphics[width=\linewidth]{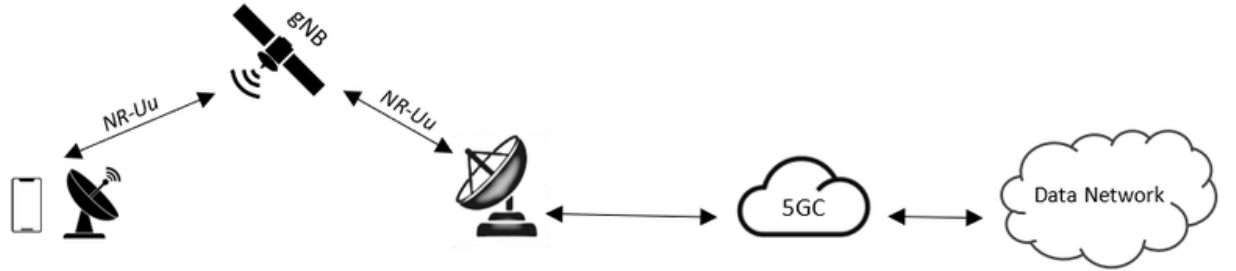}
   \caption{Fifth generation non-terrestrial network with regenerative payload.}
   \label{fig_Ng2}
\end{figure}
 
On the other hand, a regenerative payload satellite offers more advanced capabilities by performing complex signal processing tasks based on specific service requirements. This type of satellites combines all the functionalities of a gNB and provides substantial advantages. However, implementing regenerative payload systems incurs significant additional costs compared to transparent satellite systems. But their deployment is still very desirable, as the regenerative payload satellites can enable a range of new services and applications while experiencing only half of the Round-Trip Delay (RTD) because the signal processing takes place on the satellite itself. Moreover, such payloads can be promising for enabling the Inter-Satellite-Links (ISL), thus enhancing the flexibility and agility of satellite networks by enabling dynamic reconfiguration and efficient resource utilization.
However, the implementation of regenerative payload-based 5G-NTNs is still in the early stages of research, and while they bring new opportunities, they also pose unique challenges that demand extensive efforts from academia and industry. Realizing such networks will require new approaches and innovative thinking in terms of system design. Fig. \ref{fig_Ng1}-\ref{fig_Ng2} highlights an example of the deployed transparent and regenerative 5G-NTN payloads, respectively.

\section{5G-NTNs from the Protocol Stack Perspective and Challenges at the Higher Layers}

\begin{figure}[h]
    \centering
    \includegraphics[width=0.7\linewidth]{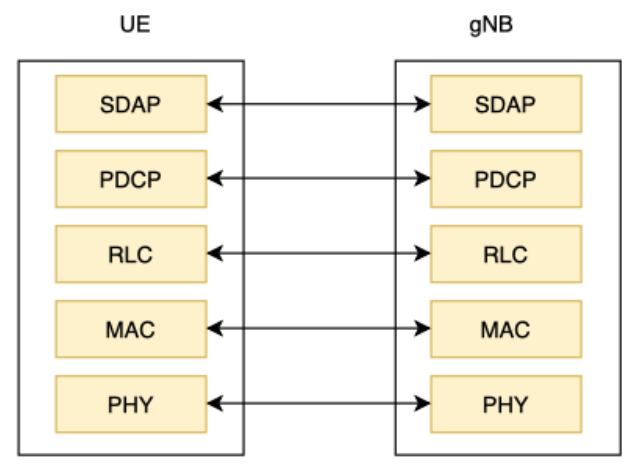}
    \caption{Fifth generation new radio user plane protocol stack.}
    \label{fig:5g-prot-stack}
\end{figure}

The NTNs radio interface consists of four primary user-plane protocol layers for secure and reliable data transfer between the UE and gNB. Such layers, which are part of the protocol stack of both the UE and gNB, are the following: 1) Physical (PHY) layer, 2) Medium Access Control (MAC) layer, 3) Radio Link Control (RLC) layer, and 4) Packet Data Convergence Protocol (PDCP) layer, as shown in Fig. \ref{fig:5g-prot-stack}. The most widely investigated layers are the PHY and MAC, for which significant contributions are available for the 5G-NTNs in  \cite{ai2019physical,topal2022physical} and \cite{ferrer2019review,wang2021joint,emmelmann2003adaptive}, respectively. However, challenges at the RLC and PDCP layers remain unexplored. The main difference between TN \cite{sheemar2022hybrid_thesis,sheemar2021hybrid,sheemar2021hybrid} and NTN is that the latter suffers from large RTDs. This leads to novel challenges to be tackled at the timing and buffer optimization levels, due to the limited memory capacity of the satellites and also to restrict the hardware cost, especially for small and low-cost regenerative architectures. We shed light on the newly emerging challenges at the higher layers in the following.

The PDCP layer, which lies on top of the RLC layer, is responsible for maintaining the integrity and confidentiality of user data. It performs header compression, duplication removal, and in-sequence delivery of user data. The PDCP layer relies on the \emph{discard timer} and the \emph{transmit buffer} at the transmitter and \emph{reordering timer} at the receiver to function. The discard timer is used to track the sojourn time of each service data unit (SDU) at the PDCP layer, which might need to be discarded from the transmission buffer when their sojourn timer exceeds the discard timer value or has been successfully received. 
The reordering timer is another very important parameter used at the receiver PDCP entity to assure that the packets are delivered to the upper layer in the correct order. Namely, when packets arrive our-of-order, the optimal value of the reordering timer should be chosen closer to the expected delay for the missing packets. The PDCP layer relies on the transmission buffer to store a copy of the data to be transmitted over the wireless link. The transmission buffer is used to ensure that the user data is transmitted in the correct sequence and that any lost packets are retransmitted, for which its copy is saved in the transmit buffer. Already larger RTDs significantly increase the number of packets to be stored in the buffer. Sub-optimally chosen discard timer values can further impact the sojourn time of the packets in the buffer, thus leading to high memory requirements to save unnecessarily an extremely large number of packets in the transmit buffer. A typical example includes the case in which the packets are correctly received on the receiver side and their acknowledgement (ACK) is sent to the transmitter. However, the ACK is lost and the transmitter instead of triggering the retransmission waits for the discard timer to expire, while other packets are being sent for each of which a copy must be saved in the buffer. Note that this can considerably enhance the delay in the system and therefore degrade the performance in terms of throughput/rate. Therefore, optimally tuned discard timer values will not only remarkably decrease the amount of memory required at the regenerative payloads but are also essential to minimize unnecessary delays in the system. This could further open the possibility to adopt NTN solutions also for the services demanding low latency.

The RLC layer is responsible for managing the reliable transmission of data and operates in three different modes. Namely, 1) RLC transparent mode (TM), 2) Unacknowledged Mode (UM), and 3) RLC Acknowledged Mode (AM). In RLC TM, the RLC layer simply passes the data received from the upper layer to the lower layer without adding any reliability mechanisms. This mode is suitable for applications that do not require high reliability, such as streaming video or audio. In RLC UM, the RLC layer sends the data without waiting for an ACK from the receiving end. The UM mode is suitable for applications that can tolerate some degree of packet loss, such as file transfers or web browsing. In RLC AM, the RLC layer uses various reliability mechanisms to ensure that all packets are successfully delivered to the receiving end. This mechanism includes segmentation, retransmission, and status reporting. The AM mode is suitable for applications that require high reliability. For the NTNs, the biggest challenge lies in the RLC AM as it performs retransmissions. Namely, in RLC AM, the receiver uses the \emph{reassembly timer} to ensure that all packets are received in the correct order and reassembled into the original data. When a packet is received out of order, the RLC layer stores it in a buffer until all the missing packets have been received. The reassembly timer is started when the first out-of-order packet is received, and it determines the maximum time that the RLC layer will wait for the missing packets to arrive. Due to large RTDs, the values of the reassembly timer should be tightly tuned, such that the packets do not unnecessarily wait before being forwarded to the higher layers. Otherwise, the throughput/rate at the higher layers can degrade.

The aforementioned challenges at the RLC and PDCP are a function of the delay. Namely, on average, the time required for the packets to be correctly received at these layers strictly depends on the position of the satellite with respect to Earth, which dictates the RTDs. Therefore, when its position changes significantly, such parameters need to be re-adapted.

Recall that the optimization of the PDCP and RLC layers in the NTNs is still an open research problem and has not yet been tackled. The solution proposed by the 3GPP so far \cite{38811, 38821} advises setting the value of the PDCP discard timer, the PDCP reassembly timer, the PDCP transmit buffer, and the RLC reassembly timer by considering the worst-case scenarios consisting of the maximum number of retransmissions allowed by the hybrid automatic repeat request (HARQ) and ARQ processes in the lower layers to achieve reliable communications. Therefore, such solutions represent an upper bound on the values of such parameters and require allocating the maximum buffer size for the PDCP layer and result in unnecessary delays for packets that require retransmission but their ACK is low. However, in practice, the effective number of retransmissions, depending on the position of the satellites, channel conditions and signal-to-noise ratio (SNR) at the PHY layer, could result to be considerably less. To this end, we present an adaptive approach to optimize the parameters of such layers by providing the tightest upper bound for the timers by estimating the effective number of retransmissions occurring in the systems. During packet loss, no ACKs are received, and the transmitter must wait until the discard timer expires. Since we propose the tightest upper bound for the discard timer value, this results in a minimum waiting time for retransmitting the lost packets. Moreover,  we remark that our proposed adaptive solutions for the timers result in a minimum amount of memory required to have no packet loss, which is very desirable for low-cost next-generation NTN solutions.

 \section{Problem Formulation and Adaptive Solutions} \label{sistema}

\subsection{PDCP Layer Optimization}

\subsubsection{\emph{PDCP Discard Timer}} Let $t_d$ denote the discard timer for the PDCP layer, which is typically configured for each data radio bearer (DRB). After receiving the service data unit (SDU) from the radio resource layer (RRC), the PDCP entity starts a timer to track its sojourn time after placing the received SDU in the transmission buffer. Either when the associated timer to each SDU exceeds the value of the discard timer $t_{d}$ or the correct reception of the PDCP SDU is confirmed, the PDCP entity should discard the PDCP SDU. 

To optimize $t_d$, we assume that its value is periodically adapted over a period of timer $T_d$ by assessing the system for a very limited duration $T_o$ in which $O$ packets are correctly received, including all the retransmissions. We propose to exploit the time stamps (TS) for each SDU received from the upper layer, which will allow the correct adaptation of $t_d$ for the NTNs. Namely, we start by setting the value of $t_d$ very large, for example as recommended by 3GPP. Typical values include $1000-1500~$ ms \cite{38811, 38821}. Let $\mathcal{T}_d = \{ t_0,t_1,....,t_O\}$ denote the set containing the indices of the TSs for the first $O$ SDUs that have been correctly received. Note that for this period packets are discarded from the transmission buffer only when the peer PDCP entity correctly receives them. However, it is possible that in timer $T_d$ some packets are correctly received at the receiver but their ACKs are lost. In such a case, the TSs of such packets can provide erroneous estimates of the effective number of retransmissions and therefore we consider discarding such samples from the set $\mathcal{T}_d$. Let $r_d$ denote the round-trip delay for the 5G-LEO satellite seen between the PHY layers. From the PDCP layer perspective, the discard timer value can be written as 
$t_d = N \;(r_d + 4 ( t_{pro}^{PDCP} + t_{pro}^{RLC} \;+\; t_{pro}^L))$, where $N \in \mathbb{N^+}$. Namely, it should be a positive integer multiple of the RTD times the number of retransmissions $N$ that effectively took place. Additionally, we have the constants $t_{pro}^{PDCP}$ and  $t_{pro}^{RLC}$ denoting the processing time delays at the PDCP layer and RLC layer, respectively, and $t_{pro}^L$ denote the total processing delay at the lower layers including MAC and PHY layer, due to for example modulation, demodulation, encoding, decoding, medium access control, flow control, compression/decompression, ciphering/deciphering of data, packet reordering, header removal etc. Finally, the scalar $4$ accounts for the path from the transmitter to the receiver up to the PDCP layer and from the receive PDCP entity to the transmit PDCP entity and therefore the delay $( t_{pro}^{PDCP} + t_{pro}^{RLC} \;+\; t_{pro}^L)$ is experienced $4$ times. Remark that such constants are inherent to the regenerative payload  and can be easily known at the systems level. Depending on the channel conditions and/or SNR, the average effective number of retransmissions required at the lower layers seen from the PDCP layer perspective may vary and thus needs to be accurately estimated. 

Consider the observation packets with their time stamps in $\mathcal{T}_d$. We aim at estimating the maximum number of retransmissions from the observation stamps so that the discard timer value can be set as tight as possible based on the effective number of retransmissions occurring in the NTN system. From the observation data in $\mathcal{T}_d$, the maximum number of retransmissions can be estimated by solving the following optimization problem.  

\begin{subequations} \label{solution_t_d_opt}
\begin{equation}
    \max_{t} \;\;\;\; \frac{t -  4(t_{pro}^{PDCP} + t_{pro}^{RLC} \;+\; t_{pro}^L )}{r_d}
\end{equation}
\begin{equation}
    \text{s.t.}  \quad\;\; t_d \in \mathcal{T}_d
\end{equation}
\end{subequations}
Such a problem can be easily solved by comparing the values of the times that the $O$ packets have stayed in the buffer available in $\mathcal{T}_d$. Let $t_d^*$ denote the optimal solution for \eqref{solution_t_d_opt}, which can be easily obtained. An estimator for the effective number of maximum retransmissions $N$ can be built as
 \begin{equation} \label{N_calc}
    N =  \lceil{\frac{t_d^* -  4(t_{pro}^{PDCP} + t_{pro}^{RLC} \;+\; t_{pro}^L)}{r_d}}\rceil
 \end{equation}
 where $\lceil \cdot \rceil$ denotes the least greatest integer.

\begin{figure}
    \centering
\includegraphics[width=7cm,height=9cm]{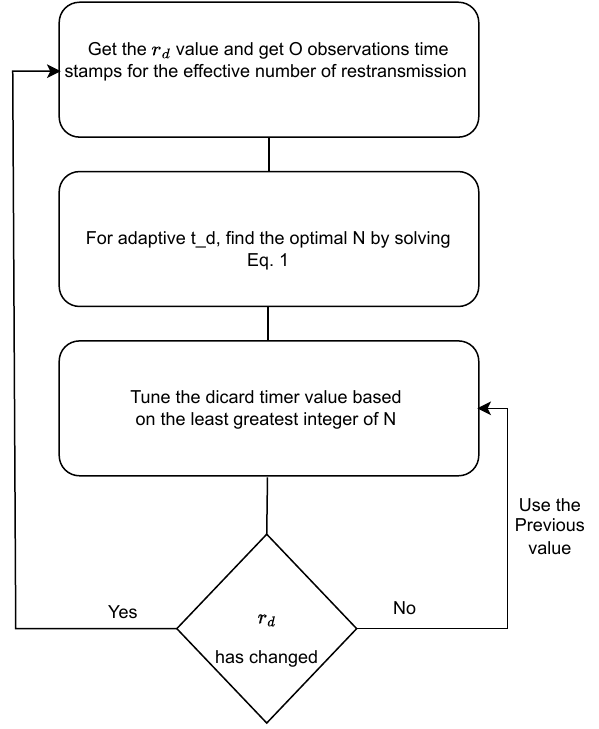}
\caption{Flowchart of the adaptive estimator of the discard timer value.}
    \label{flowchart}
\end{figure}

It should be noted that the proposed solution remains valid until time $T_D$, during which the value $r_d$ can be assumed to be constant. However, as the satellite position changes, the propagation delay also changes. Hence, it becomes necessary to adjust the discard timer value again by observing the $O$ TSs in the new time interval. This observation helps in estimating the maximum number of retransmissions in the system. The overall procedure to build an adaptive estimator for the effective maximum number of retransmissions occurring in the system is reported in Fig. \ref{flowchart}. This can value can be further used to set the optimal value of the discard timer.

\subsubsection{PDCP Transmit Buffer}
The buffer size at the PDCP layers is crucial in the upcoming generation of NTNs, particularly in compact and affordable NTNs. This is primarily because substantial propagation delays can result in a rapid escalation of the number of packets that need to be stored in the buffer while awaiting their ACKs. Consequently, this phenomenon imposes stringent demands on a large memory capacity, which can significantly increase the costs associated with NTNs.

The optimization of the transmit buffer can be based on the throughput of the system from the PDCP layer perspective at the transmit side. Furthermore, on the time required to receive the ACKs of the correct reception of the packets such that some portion of the memory can be emptied. Note that this includes the RTD and the number of retransmissions $N$ at the lower layer. Namely, let $R_p$ denote the effective rate at the PDCP layer that the transmit buffer must support to have no packet loss, denoted as the number of packets transmitted from the PDCP layer in one millisecond. The effective number of retransmission occurring in the system has been estimated when optimally tuning the discard timer value in \eqref{N_calc}. To find the optimal size of the buffer, we must calculate the total number of packets which will be stored as a copy in the transmit buffer when the transmit effective rate results to be $R_p$. It can be shown that the number of packets effectively transmitted over time while waiting for the packets' ACKs is given as 
\begin{equation} \label{opt_buff}
    \mbox{Number of Packets} = N\; ( r_d + 4 (t_{pro}^{PDCP} + t_{pro}^{RLC} \;+\; t_{pro}^L)) \;R_p.
\end{equation}
Assuming that one memory cell of the buffer stores one packet, then \eqref{opt_buff} also represents the minimum buffer size, denoted as $B^*$ in terms of memory cells required to support the throughput $R_p$ and have no packet loss in the 5G-NTN system. In case the buffer size is chosen to be small than $B^*$ is will result in packet loss at the transmitter itself, which will notify the higher layers to reduce the packet rate. Note that the ideal buffer size relies on the RTD value $r_d$. Consequently, as the regenerative payload changes its position, the RTD value and the optimal buffer size also change. Therefore, the proposed solution enables dynamic allocation of the minimum memory for 5G-NTNs to support the desired throughput/effective rate $R_p$. When the RTD is large, a significant portion of memory is allocated to store packet copies. However, when the satellite orbits closer to Earth, the RTD decreases, resulting in the release of a substantial portion of memory that can be utilized for other purposes.
 
\subsubsection{\emph{PDCP Reordering Timer}} Let $t_{r}$ denote the reordering timer, which is another key component of the PDCP layer at the PDCP receiver entity and is used to ensure that packets are delivered to the upper layers of the protocol stack in the correct order. When a packet is transmitted over the air interface, it may experience delays or be received out of order due to various factors such as congestion or interference. When a packet is received by the PDCP layer, it is assigned a sequence number (SN) based on its order of transmission. The PDCP layer checks the SN of each received packet against the expected SN, and if the received packet's SN is lower than the expected SN, it is considered an out-of-order packet and stored in the reordering buffer. The reordering timer starts when the first out-of-order packet is received and is set to a value based on the expected delay for the missing packet to arrive.  We propose to solve this challenge by exploiting the information obtained from the initial $O$ samples for which we first set the values to the maximum according to the 3GPP solution. As for different received packets, a different value of delay can be observed, and we aim at obtaining information about the maximum number of retransmissions seen at the receiver,  given the observation data. Note that on the receiver side, only half of the RTD is experienced compared to the transmit PDCP entity which must wait for the ACK. Therefore, let $r_d/2$ denote the propagation delay for the PDCP receiver. The expected delay to receive the packets at the PDCP receiver entity, including retransmissions and processing delay at the lower layers, can be written as
\begin{equation}
  t_r  =  M ( r_d/2 + 2(t_{pro}^{PDCP} + t_{pro}^{RLC} \;+\; t_{pro}^L)),
\end{equation}
The scalar $r_d/2$ takes into account the exact propagation delay from the transmitter to the receiver at the PHY layer and $M$ denote the effective number of retransmissions. We note that the number of effective retransmissions observed at the receiver may result to 
be different than $N$. This is mainly due to the fact that a packet may be received correctly at the PDCP layer, but its ACK is lost. In such a case, the transmitter will retransmit the packets for which no ACK is received, although they will be discarded on the receiver side. Therefore, $M$ should be estimated separately by the entity located on the ground. To build an estimator for the maximum number of retransmissions occurring in the system, we need to estimate and solve the following optimization problem based on the observations
 
\begin{subequations} \label{solution_t_d}
\begin{equation} \label{problem_1}
    \max_{t} \;\;\;\; \frac{t - 2 (t_{pro}^{PDCP} + t_{pro}^{RLC} \;+\; t_{pro}^L )}{r_d/2  }
\end{equation}
\begin{equation}
    \text{s.t.}  \quad\;\; t \in \mathcal{T}_r
\end{equation}
\end{subequations}
where $\mathcal{T}_r$ is a set which contains the $O$ observation TSs at the receiver, for which the packets are correctly received, including retransmission. This problem can be easily solved by comparing the ratio $\eqref{problem_1}$, with $t$ evaluated over the observation data in $\mathcal{T}_r$. Let $t_r^*$ denote the optimal solution of the problem \eqref{solution_t_d}, then the maximum expected number of retransmissions in the system is given as 
\begin{equation}
 M^* = \lceil{\frac{t_r^* -  2(t_{pro}^{PDCP} + t_{pro}^{RLC} \;+\; t_{pro}^L)}{r_d/2}}\rceil
\end{equation}


\subsection{RLC Layer Optimization}

The RLC layer with acknowledged mode (RLC AM) is responsible for providing reliable data transfer between the NTN wireless interface and the core network. It ensures the delivery of user data by performing functions such as segmentation and reassembling of data, error correction, and flow control. The RLC layer operates on top of the physical layer and provides a service to the upper layer protocols. The main challenge for the RLC layer due to large RTDs lies on the receiver side. Namely, the RLC layer uses the reassembly timer $t_{re}$ as a key parameter to ensure that all RLC protocol data units (PDUs) belonging to a specific RLC SDU are received in a timely manner. When an RLC PDU is received by the RLC layer, it is stored in a reassembly buffer. The reassembly buffer holds all RLC PDUs that belong to a specific RLC SDU until all the PDUs have been received. If an RLC PDU is lost or corrupted during transmission, the RLC layer requests a retransmission of the missing or corrupted PDU. The reassembly timer is started when the first RLC PDU belonging to an RLC SDU is received. The timer is set to a value based on the expected delay for all the PDUs belonging to the same RLC SDU to arrive. If all PDUs are not received within the timer duration, the RLC layer will discard the incomplete RLC SDU.

Since we previously estimated the effective maximum time to receive the packet which depends on the scalar $M$, at the low layer the required time is the same except for the processing time at the PDCP layer, which lies at the top of the RLC layer. Therefore, the timer $t_{re}$ can be optimally tuned as

\begin{equation}
    t_{re} = M^* (r_d/2 + 2 (t_{pro}^{RLC}  +  t_{pro}^L)) ,
\end{equation}

Note that buffer optimization is not required at the UE located on the ground as the UEs have enough memory to save a large amount of data. However, if required, a similar problem to \eqref{opt_buff} can be solved to obtain the optimal buffer size.

\section{Simulation Results}
In this section, we present simulation results to evaluate the performance of the proposed adaptive timer and buffer approach. We consider the case of one 5G LEO regenerative satellite serving one UE on Earth. We assume a round trip delay of $20$ ms for a satellite at $1200$ km
altitude and $10^\circ$ elevation angle. 
We assume that the effective number of retransmissions observed at the transmitter and receiver are the same $N=M$ and assume that it is uniformly distributed in the interval $[\mu-3,\mu + 4]$, where $\mu$ denotes the average number of retransmissions. Therefore, higher values of $\mu$ emphasize a larger number of retransmissions. The total processing delay is assumed to be $t_{pro}^{PDCP} + t_{pro}^{RLC} \;+\; t_{pro}^L = 0.5~$ms. 

For comparison, we consider the case where the discard timer value is set based on the maximum number of HARQ processes being $32$, computed according to the formula provided above. We assume that the effective rate supported by the PDCP layer is $R_p = 10 ~pck/ms$.
To yield the estimators for the effective maximum number of transmissions, we assume that a total of $1000$ packets are sent and from which the first $20$ packets are derived to estimate the effective number of retransmissions $N$ and $M$, and then the values of the discard timer $t_d$, reordering timer $t_r$ and reassembly timer $t_{re}$ are optimized. Since a similar approach is being adopted to tune the reordering timer and the reassembly timer, we consider plotting the results only for $t_d$.

\begin{figure}[t]
    \centering
    \includegraphics[width=9cm,height=5.5cm]{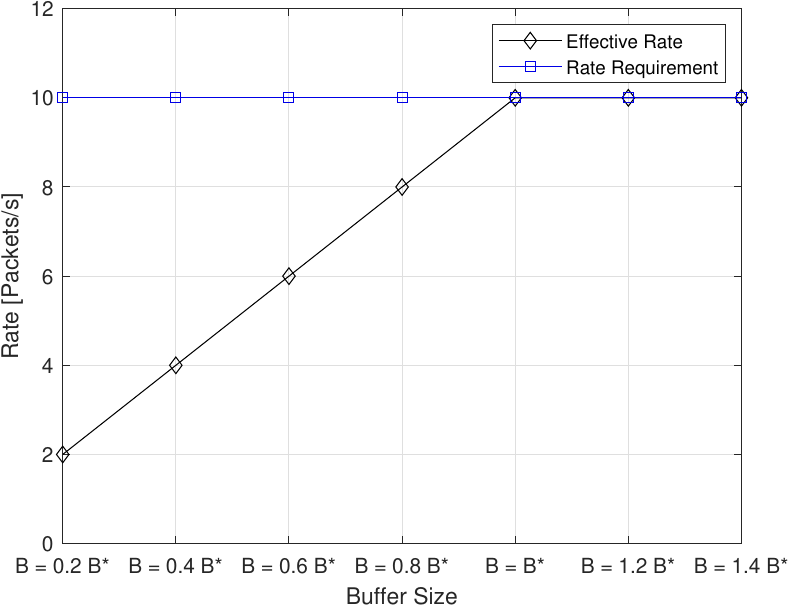}
    \caption{\footnotesize{PDCP transmission rata as a function of the buffer size.}} \label{ref_buff}
\end{figure}
 
Fig. \ref{ref_buff} highlights the throughput at the PDCP layer as a function of the buffer size that varies in the interval $[0.2 B^*, 1.4 B*]$, where $B^*$ denotes the optimal buffer size computed according to \eqref{opt_buff}. We can see that when the buffer size is chosen to be small, $B < B*$ this results in significant packet loss at the PDCP transmitter, which cannot send packets to the lower layers. That is, the effective number of transmitted packets is equal to the available memory cells in which a copy of the transmitted packets can be saved. The effective rate improves as the size of the buffer increases. We can see that the buffer size computed according to \eqref{opt_buff} results in the minimum buffer size such that the effective rate matches the rate requirement that the PDCP layer is expected to support.

\begin{figure}
    \centering \includegraphics[width=9cm,height=5.5cm]{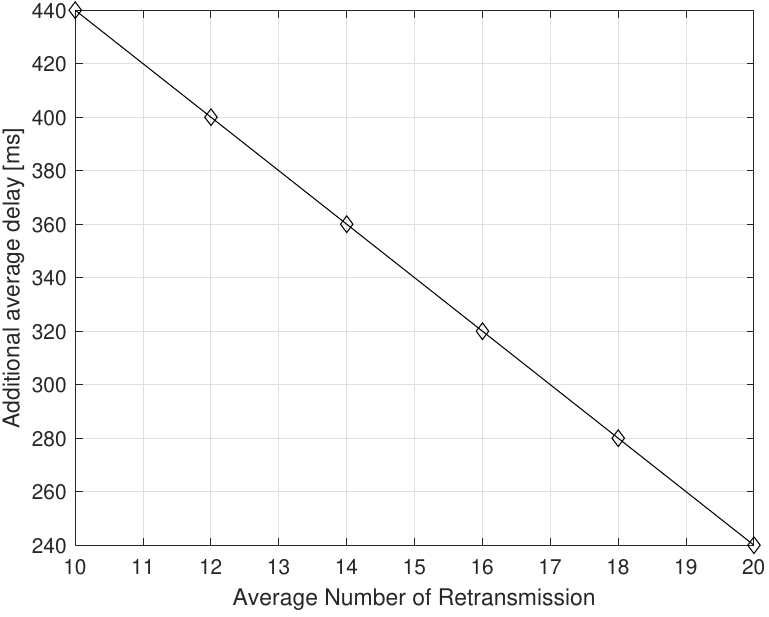}
    \caption{Average additional delay experienced by the packet which requires retransmission but its ACK is lost. }
    \label{buffer_eff_PDCP}
\end{figure}

Figure \ref{buffer_eff_PDCP} shows the average additional delay experienced for a packet which requires to be transmitted but its ACK is lost as a function of the average number of retransmission $\mu$, when the worst case for the number of retransmissions is considered, in contrast to the proposed adapted solution of this work. We can see that each packet which requires retransmission but whose ACK is lost will be retransmitted very late, especially when the effective number of retransmissions occurring in the system is small. On the other hand, when our proposed solution is adopted, the retransmission is triggered immediately as the $t_d^*$ expires based on the effective maximum number of retransmission, thus resulting in negligible additional delay.

\section{Conclusions}
The 5G-NTNs experience significant RTDs, which requires the development of innovative mechanisms for designing the higher layers of the protocol stack. This is particularly crucial for low-cost satellites with limited resources. In this study, a novel adaptive approach is introduced to optimize the discard timer, transmit buffer, reordering timer for the PDCP layer, and the reassembly timer for the RLC layer. The proposed approach accurately estimates the effective number of retransmissions in the 5G-NTN system and adjusts the timer values accordingly. In addition, it determines the optimal buffer size required to support efficient throughput. Through the results of the simulation, it is demonstrated that the proposed approach achieves significant performance improvements in terms of memory utilization efficiency. Moreover, it enables the early triggering of packet retransmissions that would otherwise unnecessarily wait for transmission due to ACK loss.

\bibliographystyle{IEEEtran}
\bibliography{main.bib}

\begin{thebibliography}{10}
\providecommand{\url}[1]{#1}
\csname url@samestyle\endcsname
\providecommand{\newblock}{\relax}
\providecommand{\bibinfo}[2]{#2}
\providecommand{\BIBentrySTDinterwordspacing}{\spaceskip=0pt\relax}
\providecommand{\BIBentryALTinterwordstretchfactor}{4}
\providecommand{\BIBentryALTinterwordspacing}{\spaceskip=\fontdimen2\font plus
\BIBentryALTinterwordstretchfactor\fontdimen3\font minus
  \fontdimen4\font\relax}
\providecommand{\BIBforeignlanguage}[2]{{%
\expandafter\ifx\csname l@#1\endcsname\relax
\typeout{** WARNING: IEEEtran.bst: No hyphenation pattern has been}%
\typeout{** loaded for the language `#1'. Using the pattern for}%
\typeout{** the default language instead.}%
\else
\language=\csname l@#1\endcsname
\fi
#2}}
\providecommand{\BIBdecl}{\relax}
\BIBdecl

\bibitem{kodheli2017integration}
O.~Kodheli, A.~Guidotti, and A.~Vanelli-Coralli, ``Integration of satellites in
  5g through leo constellations,'' in \emph{GLOBECOM 2017-2017 IEEE Global
  Communications Conference}.\hskip 1em plus 0.5em minus 0.4em\relax IEEE,
  2017, pp. 1--6.

\bibitem{kodheli2020satellite}
O.~Kodheli, E.~Lagunas, N.~Maturo, S.~K. Sharma, B.~Shankar, J.~F.~M. Montoya,
  J.~C.~M. Duncan, D.~Spano, S.~Chatzinotas, S.~Kisseleff \emph{et~al.},
  ``Satellite communications in the new space era: A survey and future
  challenges,'' \emph{IEEE Communications Surveys \& Tutorials}, vol.~23,
  no.~1, pp. 70--109, 2020.

\bibitem{giordani2020satellite}
M.~Giordani and M.~Zorzi, ``Satellite communication at millimeter waves: A key
  enabler of the 6g era,'' in \emph{2020 International Conference on Computing,
  Networking and Communications (ICNC)}.\hskip 1em plus 0.5em minus 0.4em\relax
  IEEE, 2020, pp. 383--388.

\bibitem{sheemar2021game}
C.~K. Sheemar, L.~Badia, and S.~Tomasin, ``Game-theoretic mode scheduling for
  dynamic tdd in 5g systems,'' \emph{IEEE Communications Letters}, vol.~25,
  no.~7, pp. 2425--2429, 2021.

\bibitem{sheemar2022practical}
C.~K. Sheemar, C.~K. Thomas, and D.~Slock, ``Practical hybrid beamforming for
  millimeter wave massive {MIMO} full duplex with limited dynamic range,''
  \emph{IEEE Open Journal of the Communications Society}, vol.~3, pp. 127--143,
  2022.

\bibitem{sheemar2021beamforming}
C.~K. Sheemar and D.~Slock, ``Beamforming for bidirectional {MIMO} full duplex
  under the joint sum power and per antenna power constraints,'' in
  \emph{ICASSP 2021-2021 IEEE International Conference on Acoustics, Speech and
  Signal Processing (ICASSP)}.\hskip 1em plus 0.5em minus 0.4em\relax IEEE,
  2021, pp. 4800--4804.

\bibitem{sheemar2023full}
C.~K. Sheemar, G.~C. Alexandropoulos, D.~Slock, J.~Querol, and S.~Chatzinotas,
  ``Full-duplex-enabled joint communications and sensing with reconfigurable
  intelligent surfaces,'' \emph{arXiv preprint arXiv:2306.10865}, 2023.

\bibitem{sheemar2022near}
C.~K. Sheemar, S.~Tomasin, D.~Slock, and S.~Chatzinotas, ``Near-field
  intelligent reflecting surfaces for millimeter wave {MIMO} full duplex,''
  \emph{arXiv preprint arXiv:2211.10700}, 2022.

\bibitem{hoyhtya2022sustainable}
M.~H{\"o}yhty{\"a}, S.~Boumard, A.~Yastrebova, P.~J{\"a}rvensivu, M.~Kiviranta,
  and A.~Anttonen, ``Sustainable satellite communications in the 6g era: A
  european view for multi-layer systems and space safety,'' \emph{IEEE Access},
  2022.

\bibitem{sheemar2021massive}
C.~K. Sheemar and D.~Slock, ``Massive {MIMO} mmwave full duplex relay for iab
  with limited dynamic range,'' in \emph{2021 11th IFIP International
  Conference on New Technologies, Mobility and Security (NTMS)}.\hskip 1em plus
  0.5em minus 0.4em\relax IEEE, 2021, pp. 1--5.

\bibitem{thomas2019multi}
C.~K. Thomas, C.~K. Sheemar, and D.~Slock, ``Multi-stage/hybrid bf under
  limited dynamic range for ofdm fd backhaul with mimo si nulling,'' in
  \emph{2019 16th International Symposium on Wireless Communication Systems
  (ISWCS)}.\hskip 1em plus 0.5em minus 0.4em\relax IEEE, 2019, pp. 96--101.

\bibitem{sheemar2020receiver}
C.~K. Sheemar and D.~Slock, ``Receiver design and agc optimization with self
  interference induced saturation,'' in \emph{ICASSP 2020-2020 IEEE
  International Conference on Acoustics, Speech and Signal Processing
  (ICASSP)}.\hskip 1em plus 0.5em minus 0.4em\relax IEEE, 2020, pp. 5595--5599.

\bibitem{38811}
``{3GPP TR 38.811, "3rd Generation Partnership Project; Technical Specification
  Group Radio Access Network; Study on New Radio (NR) to support non
  terrestrial networks (Release 15)"},'' 2019.

\bibitem{38821}
``{3GPP TR 38.821, "3rd Generation Partnership Project; Technical Specification
  Group Radio Access Network; Solutions for NR to support non-terrestrial
  networks (NTN) (Release 16)"},'' 2021.

\bibitem{ai2019physical}
Y.~Ai, A.~Mathur, M.~Cheffena, M.~R. Bhatnagar, and H.~Lei, ``Physical layer
  security of hybrid satellite-fso cooperative systems,'' \emph{IEEE Photonics
  Journal}, vol.~11, no.~1, pp. 1--14, 2019.

\bibitem{topal2022physical}
O.~A. Topal and G.~K. Kurt, ``Physical layer authentication for leo satellite
  constellations,'' in \emph{2022 IEEE Wireless Communications and Networking
  Conference (WCNC)}.\hskip 1em plus 0.5em minus 0.4em\relax IEEE, 2022, pp.
  1952--1957.

\bibitem{ferrer2019review}
T.~Ferrer, S.~C{\'e}spedes, and A.~Becerra, ``Review and evaluation of mac
  protocols for satellite iot systems using nanosatellites,'' \emph{Sensors},
  vol.~19, no.~8, p. 1947, 2019.

\bibitem{wang2021joint}
C.~Wang, L.~Liu, H.~Ma, and D.~Xia, ``A joint optimization scheme for hybrid
  mac layer in leo satellite supported iot,'' \emph{IEEE Internet of Things
  Journal}, vol.~8, no.~15, pp. 11\,822--11\,833, 2021.

\bibitem{emmelmann2003adaptive}
M.~Emmelmann and H.~Bischl, ``An adaptive mac layer protocol for atm-based leo
  satellite networks,'' in \emph{2003 IEEE 58th Vehicular Technology
  Conference. VTC 2003-Fall (IEEE Cat. No. 03CH37484)}, vol.~4.\hskip 1em plus
  0.5em minus 0.4em\relax IEEE, 2003, pp. 2698--2702.

\bibitem{sheemar2022hybrid_thesis}
C.~K. Sheemar, ``Hybrid beamforming techniques for massive mimo full duplex
  radio systems,'' Ph.D. dissertation, Ph. D. dissertation, EURECOM, 2022.

\bibitem{sheemar2021hybrid}
C.~K. Sheemar and D.~Slock, ``Hybrid beamforming and combining for millimeter
  wave full duplex massive mimo interference channel,'' in \emph{2021 IEEE
  Global Communications Conference (GLOBECOM)}.\hskip 1em plus 0.5em minus
  0.4em\relax IEEE, 2021, pp. 1--6.

\end{thebibliography}

\end{document}